\documentclass[twocolumn,showpacs,preprintnumbers,amsmath,amssymb]{revtex4}
\usepackage{graphicx}
\usepackage{dcolumn}
\usepackage{bm}

\begin{document}

\title{Electron transport, penetration depth and upper critical magnetic field of \\
ZrB$_{12}$ and MgB$_{2}$}
\author{V.A. Gasparov}

\author{N.S. Sidorov}
\author{I.I. Zver'kova}
\author{S.S. Khassanov}
\author{M.P. Kulakov\\}

\affiliation {Institute of Solid State Physics RAS, 142432 Chernogolovka, Moscow
District, Russian Federation \\}
\date{today}
\begin{abstract}

We report on the synthesis and measurements of the temperature dependence of
resistivity, $\rho (T)$, the penetration depth, $\lambda (T)$, and upper critical
magnetic field, $H_{c2}(T)$, for polycrystalline samples of dodecaboride
\textrm{ZrB$_{12}$} and diboride \textrm{MgB$_{2}$}. We conclude that
\textrm{ZrB$_{12}$} as well as \textrm{MgB$_{2}$} behave like simple metals in the
normal state with usual Bloch-Gr\"{u}neisen temperature dependence of resistivity and
with rather low resistive Debye temperature, $T_{R}=280~K$, for \textrm{ZrB$_{12}$}
(as compared to \textrm{MgB$_{2}$} with $T_{R}=900~K$). The $\rho (T)$ and $\lambda
(T)$ dependencies of these samples reveal a superconducting transition of
\textrm{ZrB$_{12}$} at $T_{c}=6.0~K$. Although a clear exponential $\lambda (T)$
dependence in \textrm{MgB$_{2}$} thin films and ceramic pellets was observed at low
temperatures, this dependence was almost linear for \textrm{ZrB$_{12}$} below
$T_{c}/2$. These features indicate \textit{s-wave} pairing state in
\textrm{MgB$_{2}$}, whereas a \textit{d-wave} pairing state is possible in
\textrm{ZrB$_{12}$}. A fit to the data gives a reduced energy gap $2\Delta
(0)/k_{B}T_{c}=1.6$ for \textrm{MgB$_{2}$} films and pellets, in good agreement with
published data for 3D $\pi $ - sheets of the Fermi surface. Contrary to conventional
theories we found a linear temperature dependence of $H_{c2}(T)$ ($H_{c2}(0)=0.15~T$)
for \textrm{ZrB$_{12}$}.
\end{abstract}

 \maketitle

\bigskip

\textbf{1. Introduction}

The recent discovery of superconductivity at $39K$ in magnesium diboide  \cite{1}  has
initiated a booming activity in condensed matter physics. This activity has raised
considerable interest in a search for superconductivity in other borides \cite{2}.
Unfortunately, so far none of natural candidates \textrm{MeB$_{2}$} - type diborides
of light metals (\textrm{Me = Li, Be, Al, Ca}) nor any of the large number of known
isostructural transition metal diborides (\textrm{Me = Ti, Zr, Hf, V, Ta, Cr, Mo, U})
have been found to be superconducting \cite{2}. Only in nonstoichiometric compounds
(\textrm{MoB$_{2.5}$, NbB$_{2.5}$, Mo$_{2}$B, W$_{2}B$, BeB$_{2.75}$})
superconductivity was observed \cite{3,4}. Note that the earlier speculation about
superconductivity in \textrm{TaB$_ {2}$} \cite{5} - in contradiction to other
published data \cite{2} - has been disproved by recent resistivity, susceptibility,
and the specific heat measurements \cite{6} supported by electronic structure
calculations.

These results do not seem to support the application of the old idea about
superconductivity in metallic hydrogen \cite{7} to the explanation of
superconductivity in \textrm{MgB$_{2}$} \cite{8}. In spite of this fact we would like
to discuss some aspects of this idea. In particular, it is believed that in
\textrm{MgB$_{2}$} the averaged phonon frequencies (the Debye temperature in other
words) would be very high due to the low mass of boron. The latter sharply increases
the prefactor in the McMillan formula for \textrm{T$_{c}$}. Indeed, the band structure
calculations have shown that electrons at the Fermi level are predominantly boron-like
and the superconductivity in \textrm{MgB$_{2}$} is due to graphite-type "metallic"
boron sheets \cite{8}. Furthermore, recently Eremets \textit{et al.} observed that the
semiconducting polycrystalline boron (rombohedral $\beta $ \textrm{- B$_{105}$})
transforms to a metal under high pressure and even to a superconductor at about
$160~GPa$ \cite{9}. The critical temperature \textrm{T$_{c}$} increases from $6~K$ to
$11.2~K$ at raised pressure up to $250~GPa$. This observation supports old idea that a
route for optimizing \textrm{T$_{c}$} is in preparation of the boron-rich compounds
even though this does not work yet for known borides.

In fact, the search of superconductivity in borides has a long history. Matthias et
al. discovered several superconducting cubic hexa- (\textrm{MeB$_{6}$}) and
dodecaborides (\textrm{MeB$_{12}$}) in the 60's of the last century \cite{10}. Many
other hexa- and dodecaborides (\textrm{Me=Ce, Pr, Nd, Eu, Gd, Tb, Dy, Ho, Er, Tm})
were found to be ferromagnetic or antiferromagnetic. It was suggested that the
superconductivity in \textrm{YB$_{6}$} and \textrm{ZrB$_{12}$} (having highest
\textrm{T$_{c}$} of $6.5-7.1~K$ and $6.03~K$, respectively \cite{3}), might be due to
the hypothetical cubic metallic boron. However, a much smaller isotope effect on
\textrm{T$_{c}$} for boron in comparison with \textrm{Zr} isotopic substitution
suggests that the boron in \textrm{ZrB$_{12}$} serves as inert background, and this is
\textrm{Zr} that actually is critical for superconductivity \cite{11,12}, even though
\textrm{ZrB$_{12}$} chemically contains mainly boron.

While the superconductivity in \textrm{ZrB$_{12}$} was discovered a long time ago
\cite{10}, there has been little effort devoted to the study of basic superconductive
properties of this dodecaboride. Only recently, the electron transport of solid
solutions \textrm{Zr$_{1-x}$Sc$_{x}$B$_{12}$} \cite{13} as well as the band structure
calculations of \textrm{ZrB$_{12}$} \cite{14} have been reported. Understanding the
electron transport properties of the cluster borides as well as the superconductivity
mechanism in these compounds is very important. In this paper we attempt to address
this problem. We report temperature dependent resistivity, $\rho (T)$, magnetic field
penetration depth, $\lambda (T)$, and upper critical magnetic field, $H_{c2}(T)$, for
polycrystalline samples of \textrm{ZrB$_{12}$}. Comparative data of the $\rho (T)$ and
$\lambda (T)$ in \textrm{MgB$_{2}$} thin films and pellets are also presented.

The structure of this paper is as follows. In Sec.II we report on synthesis
of \textrm{ZrB$_{12}$} and \textrm{MgB$_{2}$} and the experimental techniques.
Section III describes the electron transport in these compounds. Section IV
describes the temperature dependence of $\lambda (T)$ in thin films and
polycrystalline samples. The data on $H_{c2}(T)$ are presented in Sec.V.

\bigskip
\textbf{II. Experimental}

Under ambient conditions, dodecaboride \textrm{ZrB$_{12}$} crystallizes in the
\textit{fcc} structure (see Fig.1) of the \textrm{UB$_{12}$} type (space group
\textit{Fm3m}), $a=0.7408~nm$ \cite{15}. In this structure, the \textrm{Zr} atoms are
located at interstitial openings in the close-packed \textrm{B$_{12}$} clusters
\cite{13}. In contrast, the diborides show a phase consisting of two-dimensional
graphite-like monolayers of boron atoms with a honeycomb lattice, intercalated with
the metal monolayers \cite{2}. In our search for superconducting diboride compounds,
we observed superconductivity at $5.5~K$ in \textrm{ZrB$_{2}$} polycrystalline samples
that had a few percents amount of \textrm{ZrB$_{12}$} impurity \cite{2}. It was
recently suggested \cite{16} that this observation could be associated with
nonstoichiometry in the zirconium sub-lattice of \textrm{ZrB$_{2}$}. To resolve this
issue and to study the electron transport and basic superconducting properties of
\textrm{ZrB$_{12}$}, we successfully synthesized this compound.

\begin{figure}
\centerline{\includegraphics [width=7.5cm, height=7.5cm] {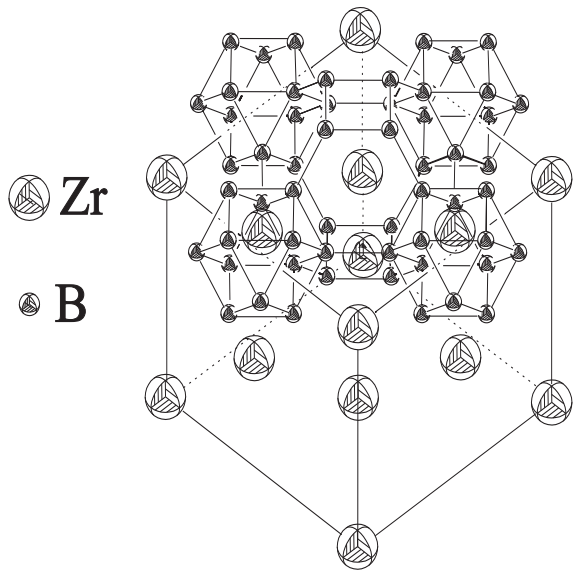}} \caption{The
lattice structure of dodecaboride \textrm{ZrB$_{12}$}. Only \textrm{B$_{12}$} clusters
on upper face of the lattice are presented for clarity.} \label{fig:1}
\end{figure}

Polycrystalline samples of \textrm{ZrB$_{12}$} were obtained by conventional
solid-state reaction. Starting materials were \textrm{Zr} metal powder ($99.99\%$
purity) and submicron amorphous boron powder ($99.9\%$ purity). These materials were
lightly mixed in appropriate amounts and pressed into pellets $10~mm$ thick and
$20~mm$ in diameter. The pellets were wrapped in a tungsten foil and baked at
$2000^{o}C$ by \textit{e-beam} heating with subsequent slow cooling to room
temperature. The process took place for two hours in a high vacuum chamber at $2\times
10^{-4}Pa$. The resulting polycrystalline pellets had over $90\%$ of the theoretical
mass density and were black in color. They demonstrated good metallic conductivity at
low temperatures. After regrinding the prepared pellets in an agate mortar, the
respective powders were reheated few times for 2 hours.

Powder X--ray diffraction pattern, obtained using \textrm{CuK}$\alpha $ radiation,
showed that the samples largely consist of desired \textrm{ZrB$_{12}$} phase (see
Fig.~2). Nevertheless, small amounts of \textrm{ZrB$_{2}$} were found to be present
and could not be eliminated by subsequent regrinding and annealing. A Rietveld
refinement of the \textrm{ZrB$_{12}$} X--ray pattern, based on the \textit{fcc}
\textrm{UB$_{12}$} type structure presented on Fig.~1, yielded lattice parameters
$a=0.7388~nm$ to be very close to the published values \cite{15}. The polycrystalline
\textrm{MgB$_{2}$} pellets have been sintered using a similar technique as outlined in
our earlier work \cite{2}. This technique is based on the reactive liquid \textrm{Mg}
infiltration of boron powder.

For this study, two highly crystalline, superconducting films of \textrm{MgB$_{2}$}
were grown on an \textit{r}-plane sapphire substrate in a two-step process. Deposition
of \textrm{B} precursor films via electron-beam evaporation was followed by
\textit{ex-situ} post annealing at $890^{o}~C$ in the presence of bulk
\textrm{MgB$_{2}$} and \textrm{Mg} vapor. Scanning electron microscopy showed dense
films with surface roughness below $5~nm$. For the measurements, we investigate films
of $500~nm$ and $700~nm$ thick, with corresponding $T_{c0}$'s of $38~K$ and $39~K$.
Details about the preparation technique are described elsewhere \cite{17}.

\begin{figure}
\centerline{\includegraphics [width=8cm, height=6cm] {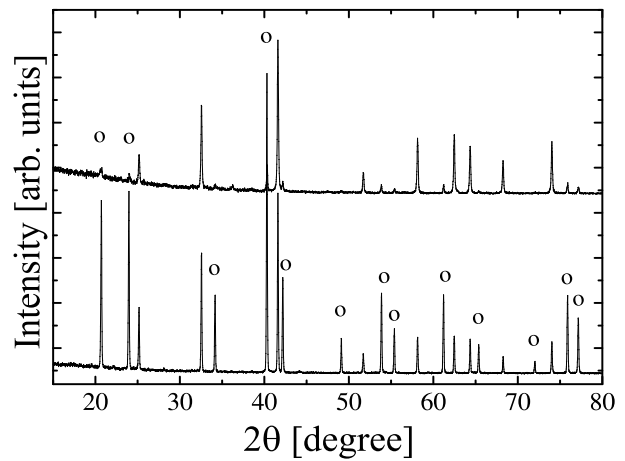}} \caption{Typical
X-ray $\theta $ -2$\theta $ scan of \textrm{ZrB$_{12}$} powders (lower curve), at room
temperature. Similar scan for \textrm{ZrB$_{2}$} pellets studied before \cite{2} is
present on upper curve. The cycles mark the X-ray reflections from \textit{fcc}
\textrm{ZrB$_{12}$}.} \label{fig:2}
\end{figure}

For the resistance measurements, we used spark erosion method to cut the
pellets into a rectangular bars with dimension of about $0.5\times 0.5\times
8~mm^{3}$. The samples were lapped with diamond paste. To remove any
deteriorated surface layers the samples were etched: \textrm{ZrB$_{12}$} in hot
nitrogen acid, \textrm{MgB$_{2}$} in $2\%$\textrm{HCl} plus water-free ethanol,
respectively. A standard four-probe \textit{ac} ($9~Hz$) method was used for
resistance measurements. Electrical contacts were made with Epotek H20E silver
epoxy. Temperature was measured with platinum (\textrm{PT-103}) and carbon
glass (\textrm{CGR-1-500}) sensors. A well-defined geometry of the samples
provided for the precise resistivity measurements.

The measurements were performed in the liquid helium variable temperature cryostat in
the temperature range between $1.1~K$ and $350~K$. Magnetic measurements of the
resistivity and penetration depth on the polycrystalline samples were carried out
using a superconducting coil in applied fields of up to $6~T$. The \textit{dc}
magnetic field was applied in the direction of the current flow. The critical
temperature measured by \textrm{RF} susceptibility \cite{2} and $\rho (T)$ was found
to be $T_{c0}=6.0~K$ and $39.0~K$ for \textrm{ZrB$_{12}$} and \textrm{MgB$_{2}$}
samples, respectively.

The $\lambda (T)$ dependence in thin films was investigated employing a single coil
mutual inductance technique. This technique, originally proposed in \cite{18} and
improved in \cite{19}, takes advantage of the well known two-coil geometry. It was
successfully used for the observation of the Berzinskii--Kosterlitz--Thouless
vortex--antivortex unbinding transition in ultrathin
\textrm{Y$_{1}$Ba$_{2}$Cu$_{3}$O$_{7-x}$} films \cite{20} as well as for the study of
the $\lambda (T)$ dependence for \textrm{MgB$_{2}$} films \cite{21}.

In particular, this radio frequency technique measures the change of inductance
$\Delta L$ of a one-layer pancake coil located in the proximity of the sample. The
coil is a part of the $LC$ circuit driven by a marginal oscillator operating at
$2-10~MHz$, or by the impedance meter (VM-508 TESLA $2-50~MHz$). The frequency
stability of this oscillator is $10~Hz$. The film is placed at small distance
($\approx 0.1~mm$) below the coil and is thermally insulated from the coil by Teflon
foil. Both sample and coil are in a vacuum, but the coil holder is thermally connected
with helium bath, while the sample holder is isolated and may be heated. During the
experiment the coil was kept at $2.5~K$, whereas the sample temperature was varied
from $2.5~K$ to $100~K$. Such design allows us to eliminate possible effects in
temperature changes in $L$ and $C$ on the measurements. The real part of the complex
mutual inductance $M$ between the film and the coil can be obtained through:

\begin{eqnarray}
\textrm{Re}\,M(T)=L_0\cdot\left(\frac{f_0^2}{f^2(T)}-1\right),
\label{eq:one}
\end{eqnarray}

Here $L_{o}$ and $f_{o}$ are the inductance and the resonant frequency of the circuit
without the sample. In the London regime, where the high frequency losses are
negligible, one can introduce the difference between temperature dependant real part
of $M$ of the coil with the sample, $ReM(T)$, and that of the coil at $T_{0}$
$ReM_{o}$. This difference is a function of the $\lambda (T)$:
\begin{eqnarray}
\Delta \textrm{Re}\,M(T)=\pi\mu_0\cdot\int_0^\infty\frac{M(q)}{1+2q\lambda
\coth(d/\lambda)}\,dq,
\label{eq:two}
\end{eqnarray}

where $M(q)$ plays the role of mutual inductance at a given wave number $q$ in the
film plane and depends on the sample-coil distance, $d$ is the sample thickness
(additional details can be found in \cite{19}). A change in $\Delta ReM(T)$ is
detected as a change of resonant frequency $f(T)$ of the oscillating signal. This
change when put into Eq.~(\ref{eq:two}) yields temperature dependent London
penetration depth $\lambda (T)$.

Measurements of $\lambda (T)$ for polycrystalline \textrm{ZrB$_{12}$} and
\textrm{MgB$_{2}$} samples were performed with a similar $LC$ technique but using a
rectangular solenoid coil into which the sample was placed. The details of this
technique are described elsewhere \cite{22}. For such arrangements, changes in the
resonant frequency of the circuit $f(T)=\omega /2\pi $ relative to that above to
$T_{c}$, $f(T_{c})$, and at minimal temperature $T_{1}$, $f(T_{1})$, are directly
related to the inductance of the probe coil and, hence, to $\lambda (T)$ by following
equation:

\begin{eqnarray}
\lambda (T)-\lambda (T_{1})=\delta \cdot \frac{f^{-2}(T)-f^{-2}(T_{1})}{f^{-2}
(T_{c})-f^{-2}(T_{1})}
\label{eq:three}
\end{eqnarray}

Here $\delta =(c^{2}\rho /2\pi \omega )^{1/2}$ is the skin depth above $T_{c}
$, which was determined from the resistivity $\rho (T)$ measurements.

\begin{figure}
\centerline{\includegraphics [width=8cm, height=6cm] {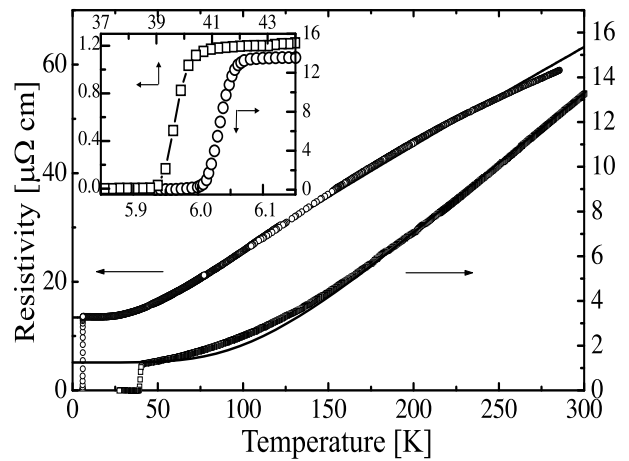}} \caption{Temperature
dependence of the resistivity $\rho (T)$ of \textrm{ZrB$_{12}$} (open circles) and
\textrm{MgB$_{2}$}(squares) polycrystalline samples. The solid lines represent BG fits
to the experimental data by Eq.~(\ref{eq:four}).} \label{fig:3}
\end{figure}

\bigskip
\textbf{III. Electron Transport}

Figure 3 shows the temperature dependence of the resistivity for \textrm{ZrB$_{12}$}
and \textrm{MgB$_{2}$} samples. Inset displays variation of $\rho (T)$ near
superconducting transition with zero resistance at $6.0~K$ (the width $\Delta
T=0.04K$) in \textrm{ZrB$_{12}$} and at $39~K$ ($\Delta T=0.7~K$) in
\textrm{MgB$_{2}$} samples. The transition is remarkably narrow for
\textrm{ZrB$_{12}$} samples, which is a clear indication of good quality samples. The
transition temperature is consistent with the previously reported values for
\textrm{ZrB$_{12}$} ($6.03~K$) \cite{10,11,12} and is comparable larger than that of
\textrm{ZrB$_{2}$} samples ($5.5~K$) \cite{2}. Despite the fact that
\textrm{ZrB$_{12}$}  contains mostly boron, its room temperature resistivity is only
four times larger than that of \textrm{MgB$_{2}$} and \textrm{ZrB$_{2}$} \cite{2},
while the residual resistivity is ten times larger. The resistivity ratio of
\textrm{ZrB$_{12}$} ($R(300~K)/R(6~K)\simeq 4$) is rather low compared to single
crystal value $10$ \cite{23}.

One can predict a nearly isotropic resistivity for \textit{fcc} \textrm{ZrB$_{12}$},
which can be described by the Bloch-Gr\"{u}neisen (BG) expression of the
electron-phonon (\textit{e-p}) scattering rate \cite{24}:

\begin{eqnarray}
\rho (t)-\rho (0)=4\rho _{1}t^{5}\int_{0}^{1/t}\frac{x^{5}e^{x}dx}{(e^{x}-1)^{2}}=
4\rho _{1}t^{5}J_{5}(1/t)
\label{eq:four}
\end{eqnarray}

Here, $\rho (0)$ is the residual resistivity, $\rho _{1}=d\rho (T)/dt$ is a slope of
$\rho (T)$ at high $T$ $(T>>T_{R})$, $t=T/T_{R}$, $T_{R}$ is the resistive Debye
temperature. As we can see from Fig.~3, the BG equation describes our data reasonably
well, indicating the importance of electron-phonon interaction for both metals. The
best fit to our data is obtained with $T_{R}=270~K$ and $900~K$ for
\textrm{ZrB$_{12}$} and \textrm{MgB$_{2}$}, respectively.

\begin{figure}
\centerline{\includegraphics [width=8cm, height=5.5cm] {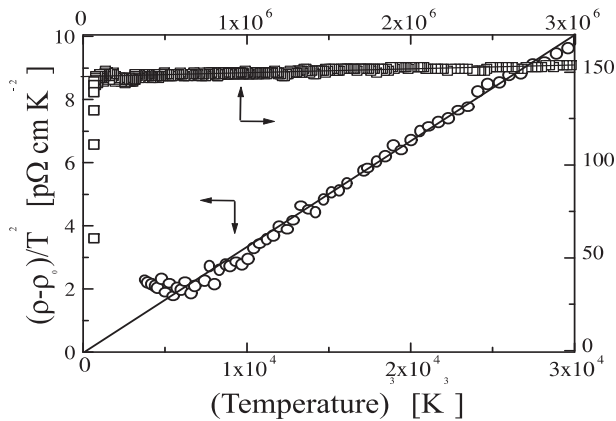}}
\caption{Temperature dependence of the reduced resistivity $[\rho (T)-\rho (0)]/T^{2}$
of \textrm{ZrB$_{12}$} (open circles) and \textrm{MgB$_{2}$} (squares) polycrystalline
samples. The solid lines describe a guide for the eye.} \label{fig:4}
\end{figure}

In contrast to \textrm{ZrB$_{12}$}, the resistivity of \textrm{MgB$_{2}$} samples does
deviate from the BG model at low temperatures Eq.~(\ref{eq:four}). This problem has
been under consideration by several groups. In particular, Putti \textit{et al.}
\cite{25} modified the BG equation introducing variable power $n$ for the
$t^{n}J_{n}(1/t)$ term in Eq.~(\ref{eq:four}). The best fit to the data was obtained
with $n=3$ which in fact ignores a small angle \textit{e-p} scattering. Recently
Sologubenko \textit{et al.} \cite{26} reported a cubic $T$ dependence in the
\textit{a,b} plane resistivity below $130~K$ in the single crystals of
\textrm{MgB$_{2}$}. This was attributed to the interband \textit{e-p} scattering in
transition metals.

We would like to stress that there are strong objections to this modified BG model:
(i) a cubic $\rho (T)$ dependence is a theoretical model for large angle \textit{e-p}
scattering and no evidence of it was observed in transition and non-transition metals;
(ii) numerous studies of the $\rho (T)$ dependence in transition metals have been
successfully described by a sum of electron-electron (\textit{e-e}), $T^{2}$, and
\textit{e-p}, $T^{5}$, contributions to the low $T$ resistivity, which may be easily
confused with a $T^{3}$ law \cite{24,27,28}; (iii) the interband $\sigma -\pi $ e-p
scattering plays no role in normal transport in the two band model for
\textrm{MgB$_{2}$} \cite{29}.

To investigate whether combination of \textit{e-e} and \textit{e-p} scattering works
for our samples we decided to add $T^{2}$ term to Eq.~(\ref{eq:four}) \cite{27,28}.
Note that the BG term is proportional to $T^{5}$ at low temperatures. Therefore,
addition of the $T^{2}$ term results in the following expression for the resistivity
$\rho (T)$:

\begin{eqnarray}
[\rho (T)-\rho (0)]/T^{2}=\alpha +\beta T^{3}
\label{eq:five}
\end{eqnarray}

\bigskip

Here $\alpha $ and $\beta $ are parameters of \textit{e-e} and \textit{e-p} scattering
terms, respectively. When plotted in $[\rho (T)-\rho (0)]/T^{2}$ vs $T^{3}$ axis such
dependence yields a straight line with slope $\beta $ and its \textit{y}-intercept
($T=0$) equal to $\alpha $. Corresponding plot of our data in Fig.~4 clearly displays
expected linear dependencies. Presence of unusually large $T^{2}$ term in
\textrm{MgB$_{2}$} data (open squares in Fig.~4) below $150~K$ is evident ($\alpha
=150~p\Omega cm/K^{2}$), whereas electron-phonon $T^{5}$ term is substantially
smaller, ($\beta =2.1\times 10^{-6}~p\Omega cm/K^{5}$). One should also note that
$\alpha $ value for \textrm{MgB$_{2}$} is almost 40 times larger than corresponding
values in transition metals such as molybdenum and tungsten ($\alpha _{Mo}=2.5~p\Omega
cm/K^{2}$ and $\alpha _{W}=1.5-4~p\Omega cm/K^{2}$ \cite{27,28}). In contrast
\textrm{ZrB$_{12}$} data display nearly zero $T^{2}$ term.

\begin{figure}
\centerline{\includegraphics [width=8cm, height=5.5cm] {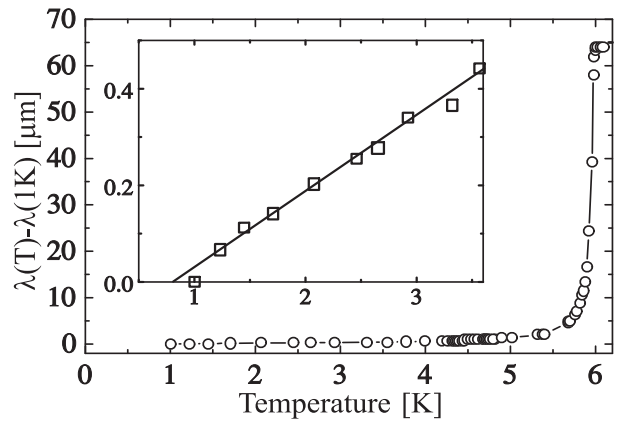}}
\caption{Temperature dependence of the penetration depth for a \textrm{ZrB$_{12}$}
sample. The solid lines describe a guide for the eye. The inset shows the data below
$3.5~K$ in an extended scale.} \label{fig:5}
\end{figure}

In general, there are many scattering processes responsible for the $T^{2}$ term in
$\rho (T)$ of metals \cite{23}. In particular, umklapp \textit{e-e} scattering
strongly contributes to this term. Furthermore normal collisions are significant in
compensated metals and in thermal resistivity \cite{28}. Borides have rather high
$T_{R}$ which depresses the \textit{e-p} scattering, so that the \textit{e-e} term is
easier to observe. Clearly there is no obvious explanation for such strong
\textit{e-e} scattering contribution in \textrm{MgB$_{2}$}. We believe additional
experiments on the more pure samples are necessary, before final conclusion about the
origin of the $T^{2}$ term in $\rho (T)$ of \textrm{MgB$_{2}$} can be drawn. Besides,
the $T^{2}$ term was recently observed in \textrm{ZrB$_{12}$} single crystal samples
with larger resistivity ratio of 10 \cite{23}. Apparently, the $T^{2}$ term is
residual resistivity dependent.

\bigskip
\textbf{IV. Penetration Depth}

Our RF technique allows us to measure the change in the penetration depth $\lambda
(T)$ \cite{22}. One should note however that there is some uncertainty in determining
the absolute values of $\lambda (T)$ for bulk samples because of error in
determination of the $f(0)$ in Eq.~(\ref{eq:two}). For this reason, we do not attempt
to determine the absolute values of $\lambda (0)$ for polycrystalline samples from
these data but rather the temperature dependent part, $\Delta \lambda (T)=\lambda
(T)-\lambda (1K)$. Fig.~5 displays the effect of superconducting transition in
\textrm{ZrB$_{12}$} on $\lambda (T)$. The striking feature of the Fig.~5 curves is the
linear temperature dependence of $\Delta \lambda (T)$ below $T_{c}/2=3K$. We should
emphasize that no frequency dependence of these data was found when oscillator
frequency was varied by two times.

In the BCS theory, the London penetration depth, $\lambda (T,l)$, is identical to the
magnetic penetration depth $\lambda (T)$ for the case of specular and diffuse surface
scattering and for negligible nonlocal effects, i.e. for $\delta (T,l)>>\xi (T,l)$
\cite{22,30}. Here $l$ - is the mean free path of carriers and $\xi$ is the coherence
length. In BCS-type superconductor (conventional \textit{s-wave} pairing) in a
clean-limit ($l>>\xi$), the $\lambda (T)$ has an exponentially vanishing temperature
dependence bellow $T_{c}/2$ (where $\Delta (T)$ is almost constant) [30]:

\begin{eqnarray}
\lambda (T)=\lambda (0)\cdot \lbrack 1+\sqrt{\pi \frac{\Delta (0)}{2k_{B}T}}\times
\exp (-\frac{\Delta (0)}{k_{B}T})],
\label{eq:six}
\end{eqnarray}

Here $\Delta (0)$ is the value of the energy gap and $\lambda (0)$ is a
magnetic penetration depth at zero temperature.

At the same time, the unconventional \textit{d-wave} pairing symmetry causes the
energy gap to be suppressed along node lines on the Fermi surface. The latter results
in a linear dependence of $\lambda (T)-\lambda (0)\propto T$ at low temperatures. Such
a linear $T$ dependence of the $\lambda (T)$ was currently used as a fingerprint of
\textit{d-wave} symmetry for Cooper pairs in cuprate superconductors \cite{31,32}.
From this point, one could argue that the linear $\lambda (T)$ dependence in
\textrm{ZrB$_{12}$} (Fig.~5) may be considered as indication of \textit{d-wave}
symmetry of Cooper pairs condensate.

Recently, however, thermodynamic arguments were suggested \cite{33}, that a strictly
linear $T$ dependence of $\lambda (T)$ at low temperatures violates the third law of
thermodynamics, since it produces the non vanished entropy in the zero temperature
limits. Therefore one should expect a deviation from the linear $T$ dependence of
$\lambda (T)$ at very low temperatures. Indeed, recent experiments in cuprates
indicate deviation from the linearity of $\lambda (T)$  from current carrying zero
energy surface Andreev bound states \cite{34}. We believe that further experiments on
single crystals of \textrm{ZrB$_{12}$} are necessary to confirm actual character of
the $\lambda (T)$ behavior below $1.0~K$. Such experiments are in progress and may
shed light on the nature of pairing state in this dodecaboride.

\begin{figure}
\centerline{\includegraphics [width=8cm, height=5.7cm] {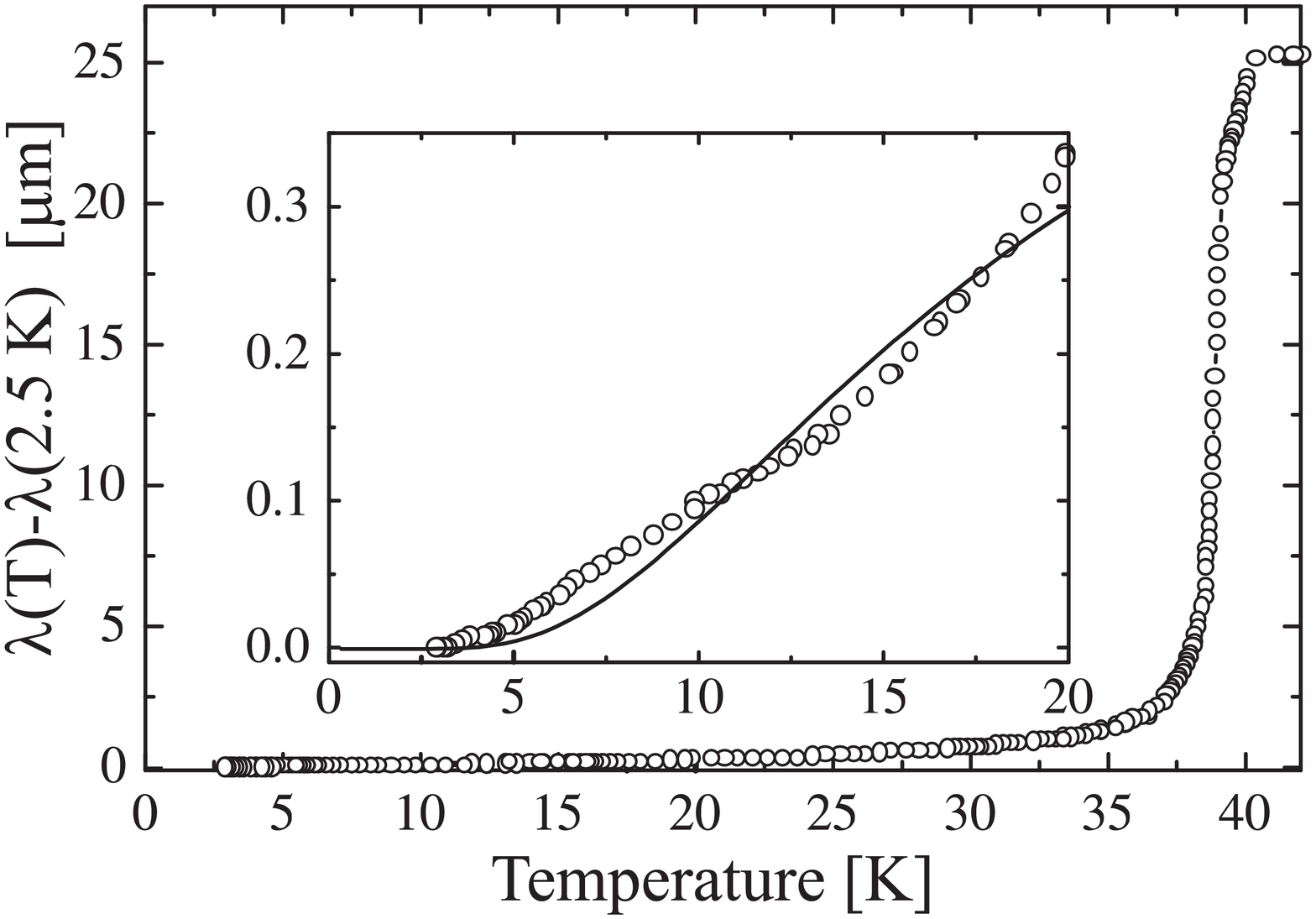}}
\caption{Temperature variation of the magnetic penetration depth, $\lambda (T)$, for
\textrm{MgB$_{2}$} sample up to $T_{c}$. The inset shows the data below $T_{c}/2$ on
an extended scale. The solid line represents the single gap exponential fit for
$\Delta (0)=2.73~meV$.} \label{fig:6}
\end{figure}

Fig.~6 displays the change in $\lambda (T)$ in \textrm{MgB$_{2}$} polycrystalline
sample. These measurements were done on samples freely placed in rectangular solenoid
coil forming $LC$ circuit which was kept at $2.5~K$. In Fig.~7 we show the temperature
variation of $\lambda (T)$ for the best \textrm{MgB$_{2}$} film as determined from
one-coil technique and inversion procedure by use of Eq.~(\ref{eq:two}). A particular
feature of these figures is a very similar exponential $T$ dependence at low
temperatures for both film and polycrystalline samples. We used conventional
\textit{s-wave} approach Eq.~(\ref{eq:six}) to fit these data. In both case we observe
satisfactory if not perfect agreement between the fits and low temperature data for
thin films. Our fitting parameters (superconducting gap value at 0K) are $2.8~meV$ and
$2.73~meV$, for film and polycrystalline samples respectively. Corresponding reduced
gap $2\Delta (0)/k_{B}T_{c}$ for these samples was found to be $1.64$ and $1.62$.

Several recent reports on $\lambda (T)$ measurements \cite{21,35} in
\textrm{MgB$_{2}$} provide strong evidence for a predominately exponential temperature
dependence of $\lambda (T)$ at low temperatures, which is consistent with our
observation. The reduced gap deduced from exponential fits to the data was found to be
$1.42$ \cite{35} and $2.3$ \cite{21} for single crystals and thin films respectively.
These values, as well as the value we obtained from our data, are significantly
smaller than the BCS weak coupling value $2\Delta (0)/k_{B}T_{c}=3.52$. Several other
groups have claimed that $\lambda (T)$ in \textrm{MgB$_{2}$} does follow a power law
or even linear $T$ dependence \cite{36}. The possible reason for this discrepancy is
that previous studies have been limited to temperatures above $4~K$, whereas $\lambda
(T)$ shows a clear signature of exponential behavior only below $7~K$ (see Fig.~6 and
Fig.~7). Another problem may arise in use of non etched samples where the damaged
surface layer or the proximity effect, associated with the presence of a metallic
\textrm{Mg} overlayer \cite{17}, may significantly complicate the use of the surface
sensitive techniques.

We would like to emphasize that our value of the superconducting gap at low $T$ are in
the range of values for 3D $\pi $ - bands obtained by point-contact spectroscopy on
\textrm{MgB$_{2}$} single crystals ($\Delta _{\sigma }=7.1~meV$ and $\Delta _{\pi
}=2.9~meV$ for the $\sigma $ and $\pi $ bands, respectively) \cite{37}. Our date also
agree with the theoretical values predicted by the two-band model \cite{38}. Analysis
of overall temperature dependence of $\lambda (T)$ dependence by aid of a two band
phenomenological model \cite{39} is in progress now and will be published elsewhere.
The essential property of this paper is comparison a \textrm{ZrB$_{12}$} and
\textrm{MgB$_{2}$} low temperature data, where $\lambda (T)$ dependence has completely
different behavior.

\begin{figure}
\centerline{\includegraphics [width=8cm, height=5.5cm] {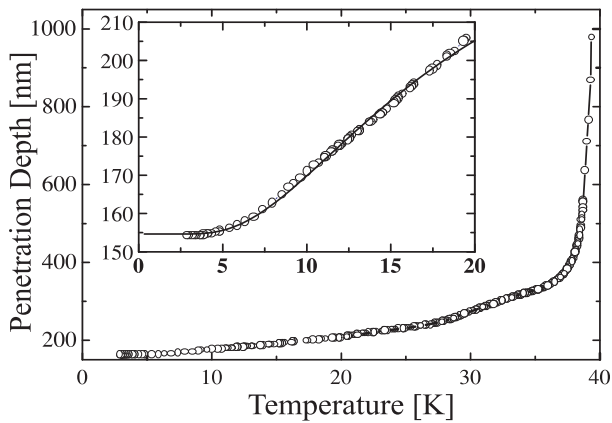}} \caption{
Temperature variation of $\lambda (T)$ up to $T_{c}$ for an \textrm{MgB$_{2}$} thin
film on \textrm{Al$_{2}$O$_{3}$}. The inset shows the data below $T_{c}/2$ on extended
scale. The solid line represents the single gap exponential fit for $\Delta
(0)=2.8meV$.} \label{fig:7}
\end{figure}

\bigskip
\textbf{V. Upper Critical Magnetic Field}

We now turn to the data on electronic transport in magnetic field. Fig.~8 presents the
magnetic field dependent electrical resistivity data for a \textrm{ZrB$_{12}$}
polycrystalline samples at various temperatures. Two features are clearly seen: (i)
magnetic field shifts the superconducting transition to lower temperatures, (ii) there
is a very small magnetoresistivity in the normal state. We extracted completion upper
critical magnetic field $H_{c2}$ by extending the maximal derivative $d\rho /dH$ line
(dashed line in Fig.~8) up to the normal state level. The crossing point of this line
and normal state resistivity gave us the value of $H_{c2}$ at various temperatures as
indicated by arrow in Fig.~8. Despite a clear broadening at higher fields, such onset
of resistive transition remains well defined. One should note however that the
resistance may not be an intrinsic property and may be related to the poor grain
connection in our polycrystalline samples. Therefore, to get a better test for the
onset of the superconducting transition we measured \textrm{RF} susceptibility.

Figure 9 shows a plot of the temperature dependence of the resonant frequency of our
$LC$ circuit, $f(T)$, as a function of longitudinal magnetic field. The set-up of the
sample arrangement is shown on an inset in the Fig.~9. Changes in the resonant
frequency are directly proportional to the \textrm{RF} susceptibility of the sample.
To deduce the $H_{c2}(T)$, we used a straight-line fit representing the maximum of
derivative $df/dH$ (dashed line in Fig.~9). This straight line was extended up to
normal state frequency values. We defined $H_{c2}$ as an crossing point of this line
with normal state frequency $f(T)$. As we can see from Fig.~9, this point is very
close to the onset point of $f(T)$ in this plot, that makes determination of
$H_{c2}(T)$ more reliable.

\begin{figure}
\centerline{\includegraphics [width=8cm, height=6cm] {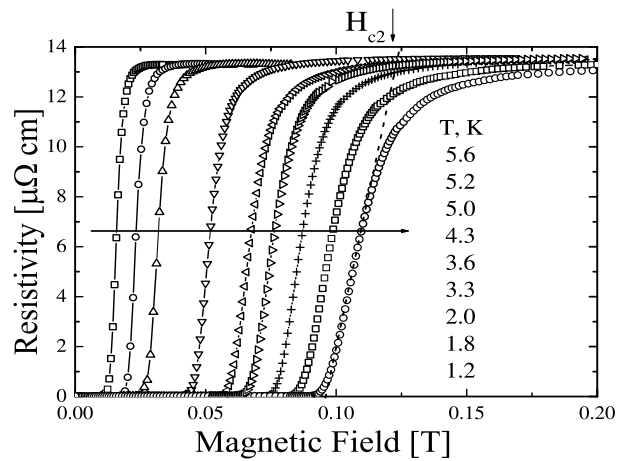}} \caption{Magnetic
field variation of the resistivity $\rho (T)$ in linear scale for a
\textrm{ZrB$_{12}$} sample. The solid lines describe a guide for the eye and the
dashed line describes how the resistive transition $H_{c2}$ has been established.}
\label{fig:8}
\end{figure}

Figure 10 presents the $H_{c2}(T)$ data that we deduced from these two techniques. A
remarkable feature of this plot is the near linear increase of $H_{c2}$ with
decreasing temperature for both data with no evidence of saturation down to $1.1K$. To
obtain the value of $H_{c2}(0)$ from our \textrm{RF} data we assumed that
$H_{c2}(0)=0.71T_{c}dH_{c2}/dT$ \cite{40} at zero temperature. This assumption yields
$H_{c2}(0)=0.11~T$, which is substantially smaller than the extrapolated value
$0.15~T$. Nevertheless, we used this number to obtain the coherence length $\xi (0)$,
by employing the relations $H_{c2}(0)=\phi _{0}/2\pi \xi ^{2}(0)$. The latter yields
$\xi (0)=60.3~nm$, the value which is substantially larger than a few angstroms
coherence length of high-Tc superconductors. The accuracy of our measurements of
$\lambda (T)$ in \textrm{ZrB$_{12}$} did not allow us to determine the absolute values
of $\lambda (0)$. Therefore, the Ginzburg-Landau parameter, $\kappa =\lambda /\xi $
could not be determined from the measurements.

Taken as a whole, the temperature dependence of $H_{c2}(T)$ for \textrm{ZrB$_{12}$} is
very similar to that found in \textrm{MgB$_{2}$} \cite{41,42} and \textrm{BaNbO$_{x}$}
\cite{43} compounds. Unlike conventional BCS theory \cite{40}, $H_{c2}(T)$ is linear
over an extended region of temperatures with no evidence of saturation at low $T$.
Although the origin of this feature is not completely understood, similar linear
$H_{c2}(T)$ dependence have been observed in other high borides and oxide compounds
\cite{41,42,43}.

\begin{figure}
\centerline{\includegraphics [width=8cm, height=5.5cm] {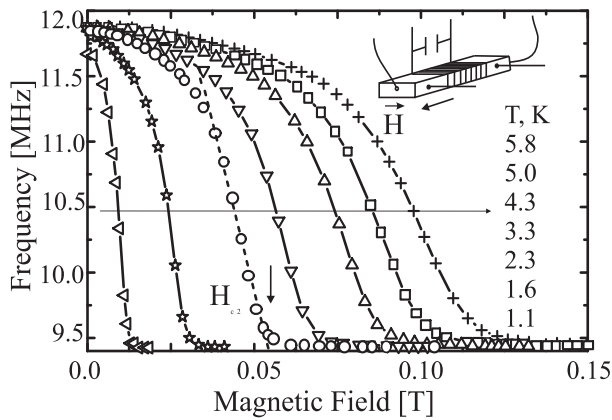}} \caption{ Magnetic
field variation of the resonant frequency of $LC$ circuit for an \textrm{ZrB$_{12}$}
sample, at different temperatures. The solid lines are guide to the eye and the dashed
line describes a linear extrapolation of the \textrm{RF} data used for $H_{c2}(T)$
determination.} \label{fig:9}
\end{figure}

\bigskip
\textbf{VI. Conclusions}

We successfully performed syntheses of polycrystalline samples of dodecaboride
\textrm{ZrB$_{12}$} and diboride \textrm{MgB$_{2}$}. We systematically studied the
temperature dependence of resistivity, $\rho (T)$, the magnetic penetration depth,
$\lambda (T)$, and upper critical magnetic field, $H_{c2}(T)$, in these compounds. The
electron transport and superconducting properties have been compared with aim to shade
light on the origin of superconductivity in borides. Although a standard
Bloch-Gr\"{u}neisen expression describes the resistivity data fairly well in
\textrm{ZrB$_{12}$}, better fit was obtained by adding electron-electron scattering
$T^{2}$ term in $\rho (T)$ of \textrm{MgB$_{2}$}. This square term dominates the $\rho
(T)$ dependence below $150~K$ in \textrm{MgB$_{2}$}, though is almost zero for
\textrm{ZrB$_{12}$}.

\begin{figure}
\centerline{\includegraphics [width=8cm, height=5.5cm] {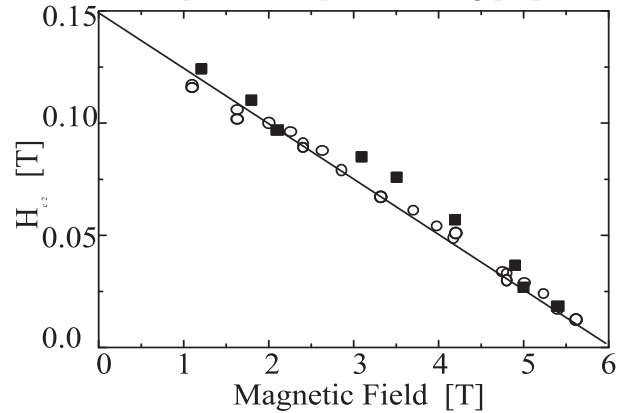}}
\caption{Temperature variation of upper critical magnetic field of \textrm{ZrB$_{12}$}
sample. Different symbols represent the data determined from $\rho (H)$ (squares) and
$f(H)$ (circles) data as described in text.} \label{fig:10}
\end{figure}

The temperature dependence of $\lambda (T)$ of both polycrystalline and thin film
\textrm{MgB$_{2}$} samples is well described by a \textit{s-wave} behavior of the
order parameter symmetry. Our value of the reduced superconducting gap in
\textrm{MgB$_{2}$} samples ($2\Delta (0)/k_{B}T_{c}=1.6$) is significantly smaller
than the weak coupling BCS value. However this value is in a very good agreement with
other direct probe measurements of smaller gap - on the $\pi $ - sheets of the Fermi
surface. At the same time, we find that $\lambda (T)$ in \textrm{ZrB$_{12}$} has
linear temperature dependence over an extended region of $T$. This feature may be
indicative of the \textit{d-wave} pairing. We find that the $H_{c2}(T)$ deduced from
\textrm{RF} data is almost the same as an that obtained from the resistive data. Both
techniques demonstrate unconventional linear temperature dependence of the
$H_{c2}(T)$, with a considerably lower value of $H_{c2}(0)=0.15~T$. We believe that
these observations are clear indicators of the unconventional behavior of electron
transport and superconducting properties of dodecaboride \textrm{ZrB$_{12}$}.

\begin{acknowledgments}
Very useful discussion with V.F. Gantmakher, A.A. Golubov, A. Junod, H.
Hilgenkamp, R. Huguenin, help in samples preparation from V.V. Lomejko and
help in text preparation of L.V. Gasparov, are gratefully acknowledged. Many
thanks to H.M. Christen, H.-Y. Zhai, M. Paranthaman, and D.H. Lowndes for
preparation of excellent \textrm{MgB$_{2}$} films. This work was
supported by the Russian Scientific Programs: Superconductivity of
Mesoscopic and Highly Correlated Systems (Volna 4G); Synthesis of Fullerens
and Other Atomic Clusters (No.541-028); Surface Atomic Structures
(No.4.10.99), Russian Ministry of Industry, Science and Technology
(MSh-2169.2003.2), RFBR (No.02-02-16874-a) and by the INTAS (No.01-0617).
\end{acknowledgments}

\end{document}